\documentstyle[12pt,fleqn]{article}
\def\beeq{\begin{equation}}
\def\eneq{\end{equation}}
\def\beeqa{\begin{eqnarray}}
\def\eneqa{\end{eqnarray}}

\addtocounter{section}{-1}
\begin{document}

\begin{center}

\vspace{2cm}

{\large {\bf {
New type of antiferromagnetic state in stacked nanographite
} } }

\vspace{1cm}

{\rm Kikuo Harigaya\footnote[1]{E-mail address: 
\verb+harigaya@etl.go.jp+; URL: 
\verb+http://www.etl.go.jp/+\~{}\verb+harigaya/+;
Address after April 2001: National Institute of
Advanced Industrial Science and Technology (AIST), 
Tsukuba 305, Japan}}

\vspace{1cm}

{\sl Physical Science Division,
Electrotechnical Laboratory,\\ 
Umezono 1-1-4, Tsukuba 305-8568, Japan}\footnote[2]{Corresponding address}\\
{\sl National Institute of Materials and Chemical Research,\\ 
Higashi 1-1, Tsukuba 305-8565, Japan}\\
{\sl Kanazawa Institute of Technology,\\
Ohgigaoka 7-1, Nonoichi 921-8501, Japan}

\end{center}

\vspace{1cm}

\noindent
{\bf Abstract}\\
Nanographite systems, where graphene sheets of the orders of 
the nanometer size are stacked, show novel magnetic properties, 
such as, spin-glass like behaviors and the change of ESR line 
widths in the course of gas adsorptions.  We theoretically 
investigate stacking effects in the zigzag nanographite sheets 
by using a tight binding model with the Hubbard-like onsite 
interactions.  We find a remarkable difference in the magnetic 
properties between the simple A-A and A-B type stackings. 
For the simple stacking, there are not magnetic solutions.
For the A-B stacking, we find antiferromagnetic solutions 
for strong onsite repulsions.  The local magnetic moments 
tend to exist at the edge sites in each layer due to the 
large amplitude of wavefunctions at these sites. 
We study variations between the A-A and A-B stackings
to find that the magnetism between the two stackings
experiences the first order phase transition.  The effect
of the interlayer distance is also discussed.


\pagebreak

\section{Introduction}

Nanographite systems, where graphene sheets of the orders of 
the nanometer size are stacked, show novel magnetic properties, 
such as, spin-glass like behaviors [1], and the change of 
ESR line widths while gas adsorptions [2]. Recently, it 
has been found [3] that interlayer magnetic interactions 
increase with the decrease of the interlayer distance 
while water molecules are attached physically.  The magnetic
susceptibility decreases due to the enhanced magnetic 
coupling between layers.  Here, the change of the inter-layer 
interactions has been anticipated experimentally, but 
theoretical studies have not been reported yet.

In this paper, we theoretically consider the stacking 
effects in the zigzag nanographite sheets [4-6] by using a tight 
binding model with the Hubbard-like onsite interactions $U$. 
In the papers [4-6], the one dimensional graphite ribbons
have been investigated.  In this paper, we assume that
each graphite sheet has a hexagonal shape with zigzag edges.
Such the shape geometry has been used in the semi-empirical 
study of fluorine doped graphite nanoclusters [7], too.
The two stacking types, namely the A-A and A-B types, shown 
in Fig. 1 are of particular interests.  The circles in 
Fig. 1 (a) (namely, nanographite {\bf a}) show the sites 
where the interlayer distance between carbon atoms is shortest.  
The zigzag edge sites have very weak interactions with 
the neighboring layers.  On the other hand, in Fig. 1 (b) 
(nanographite {\bf b}), all of the adjacent carbon atoms 
interact within each layer and with neighboring layers.

In studying the magnetic properties of the stacked systems,
it is interesting to look at the continuous change between
the nanographites {\bf a} and {\bf b}.  We will move the
first layer of Fig. 1 (b) to the upper direction.  When the 
relative shift $d=0$, the geometry is of Fig. 1 (b).  As
$d$ becomes larger, the system changes from the A-A stacking
[Fig. 1 (b)] to the A-B stacking [Fig. 1 (a)].  When $d = a$ 
($a$ is the bond length in each layer), the system has the 
geometry of Fig. 1 (a).  In increasing $d$, the edge sites feel
weaker interactions from neighboring layers.  Such the
changes will give rise to variations in magnetic properties.
The main purpose of this paper is to investigate
changes of magnetism in stacked nanographite systems.
We also discuss how magnetic properties change while
interlayer distance $R$ varies.

The main finding of this paper is a remarkable difference 
in the magnetic properties between the simple A-A and A-B 
stackings.  For the simple stacking, we have not found
magnetic solutions, because the presence of local magnetic 
moments is suppressed at carbons.  For the A-B stacking, 
we have found antiferromagnetic solutions for $U>2t$, $t$ 
being the hopping integral in a layer. The local magnetic 
moments tend to exist at the edge sites in each layer due 
to the large amplitude of wavefunctions at these sites. 
Therefore, the A-B type stacking is favorable in order that 
the exotic magnetism is observed in nanographite systems.

We also study continuous changes between the A-A and A-B 
stackings.  We will find that the magnetism between the two 
stackings experiences the first order phase transition.  
The interlayer magnetic coupling becomes stronger when 
the interlayer distance becomes shorter.  
Relations with experiments are discussed extensively.

In Sec II, we explain our model.  Sections III and IV are 
devoted to the total magnetic moment per layer, and
the local magnetic polarization per site, respectively.
In Sec. V, we discuss the local density of states
at the edge carbon atoms.  This paper is closed
with summary in Sec. VI.

\section{Model}

We study the following model with hopping integrals between
orbitals of carbon atoms and onsite strong repulsions of
the Hubbard type:
\beeqa
H &=& -t \sum_{\langle i,j \rangle: {\rm intralayer}} \sum_\sigma
(c_{i,\sigma}^\dagger c_{j,\sigma} + {\rm h.c.}) \nonumber \\
&-& \sum_{(i,j): {\rm interlayer}} \sum_\sigma
\beta (r_{i,j})
(c_{i,\sigma}^\dagger c_{j,\sigma} + {\rm h.c.}) \nonumber \\
&+& U \sum_i n_{i,\uparrow} n_{i,\downarrow},
\eneqa
where $n_i = c_{i,\sigma}^\dagger c_{i,\sigma}$ for 
$\sigma = \uparrow$ and $\downarrow$; $c_{i,\sigma}$ is
an annihilation operator of an electron at the $i$th site
with spin $\sigma$; the sum of the first line is taken
over the nearest neighbor pairs $\langle i,j \rangle$
in a single layer of the nanographite; the sum of the
second line is taken over pairs of the sites $(i,j)$
in neighboring layers; the function $\beta (r)$ is given 
by Eq. (2) (shown below); $r_{i,j}$ indicates the distance 
between the $i$th and $j$th carbon atoms; and the last 
term of the hamiltonian is the strong onsite repulsion 
with the strength $U$.

The interlayer interaction is taken into account with
the functional form:
\beeq
\beta(r) = A {\rm exp} (-r/\zeta)
\eneq
where $r$ is the distance between carbon atoms, $A = 5.21t$,
and $\zeta = 0.86$\AA.  The magnitude, $\beta(r=3.40$\AA$)
= 0.1t$, is a typical value for the interlayer interaction
strength in the tight binding model for A-B stacked graphite 
layers [8]:  the explicit value is about 0.35 - 0.39 eV, 
and $t \sim 3$eV gives the interaction strength about $0.1t$.
The similar semi-empirical forms with the exponential
dependence on the distance have been used for
C$_{60}$-polymers [9], multi-wall carbon nanotubes [10],
and many chains of conjugated polymers [11], in literatures.
The interchain interactions in conjugated polymers are
of the order $0.1t$ at most [11,12] also, and such the exponential 
dependence well describes the weak hopping interactions 
between the adjacent and nearer carbon atoms.

The finite size system is diagonalized numerically,
and we obtain two kinds of solutions.  One of them
is a nonmagnetic solution where up and down spin 
electrons are not polarized in each layer.  This 
kind of solutions can be found in weak $U$ cases.  
The other kind of solutions is an antiferromagnetic 
solution, where the number of up spin electrons is 
larger than that of down spin electrons in the first 
layer, the number of down spin electrons is larger 
than that of the up spin electrons in the second 
layer, and so on.  This kind of solution is realized 
in strong $U$ regions.  There will be cases of 
incommensurate spin density waves, but we have not 
obtained such kinds of solutions by choosing initial 
magnetic ordered states, which are commensurate 
with the one dimensional lattice in the stacking 
direction, at the first stage of the numerical 
iteration process.  The present author has discussed
the antiferromagnetism in C$_{60}$ polymers [13].  
The same technique used in Ref. [13] is effective 
in this paper, too.

The parameters are changed within $0 \leq d \leq 1.0a$
(a is the bond length in a layer), $3.0{\rm \AA} \leq R 
\leq 3.4{\rm \AA}$, and $0 \leq U \leq 3t$.  
All of the quantities of the energy dimension are
reported using the unit $t$ ($\sim 3.0$ eV).

\section{Magnetic moment per layer}

In this section, we report the total magnetic moment
in a layer.  We discuss dependences on several model
parameters: namely, the Coulomb interaction strength
$U$, the relative shift between neighboring layers $d$,
and the perpendicular distance between layers $R$.

First, we discuss the effects of the
relative motion of one layer with respect to the 
neighboring one.  Figure 2 shows the absolute value of 
the total magnetic moment per layer with changing $d$ 
and $U$.  The interlayer distance is $R=3.4$\AA.  
See the figure caption for the parameter $U$.  There is 
no magnetic moment for $0 \leq U \leq 2.1t$ for all 
the $d$.  The finite magnetization appears with the 
antiferromagnetic ordering in the one dimensional 
direction for $U \geq 2.2t$ and $0.5a < d \leq 1.0a$.
The appearance of the magnetization with respect
to increasing $U$ is continuous, and this is the
second order phase transition.  The magnetic moment
increases as a function of $U$.  On the other hand,
there is not any magnetic ordering for all the $U$
in the region $0 \leq d \leq 0.5a$.  For larger $U$,
the finite magnetization appears suddenly with
the discontinuity at $d = 0.5a$.  Therefore,
the variation with respect to $d$ is the first
order phase transition.

The remarkable difference on the magnetic properties
between small and larger $d$ is manly due to the 
change of the stacking properties.  When $d \sim 0$,
the system is near the A-A stacking.  When $d \sim 1.0a$,
it is near the A-B stacking.  Such the difference
will be discussed taking into account of the local magnetic
polarization in the next section.

Next, we look at the effects of the interlayer distance
$R$.  Figure 3 displays the absolute magnitude of the 
total magnetic moment per layer with changing $d$ and $R$.  
The strength of the Coulomb interaction is $U=2.5t$.  
The values of $R$ are $R=3.4$\AA (filled squares), 3.2\AA
(open squares), and 3.0\AA (filled circles), 
respectively.  In the experiment [14], the change
of the interlayer distance from 3.4\AA to 3.8\AA
with desorption of the water molecules has been 
found.  The chemical pressure from the adsorbed
water becomes loose, and this gives rise to the
change of the interlayer distance.  This is an example
of experiments which are related with the present
calculation (even though the model parameter value
is somewhat different and this is a minor problem).
In decreasing $R$, the interlayer hopping interactions
become stronger.  This gives rise to the increase
of the total magnetic moment per layer where
the finite antiferromagnetic order is present.
The change of magnetization between the small
and larger $d$ is always discontinuous, and the
first order phase transition occurs.  The exponential
dependence $\beta(r)$ could give change of the
magnetic moment over several orders of magnitude
in general.  But, we have limited the change
of $R$ in a certain region.  Even though the 
parameter region is quite small, we can note
that the magnetization increases by several
times with only the slight decrease of the 
interlayer distance presumably by a static
pressure or by some chemical pressure effects.
Therefore, we conclude that the variation of 
magnetic properties by some modification of 
the stackings is a quite controllable one.

\section{Local magnetic polarization in a layer}

Here, we report the magnetic moment per site
in a layer, particularly paying attention to 
the edge sites.

Figure 4 shows the local magnetic moment at the 
edge sites A, B, and C (displayed in Fig. 1 (a))
with changing the relative motion $d$ between layers.
These sites have weak interactions with neighboring
layers when the system is near the A-B stacking
$d \sim 1.0a$.  In Fig. 4, the other parameters are 
$U=2.5t$ and $R=3.4$\AA.  The filled squares, open 
squares, and filled circles show the results at sites 
A, B, and C, respectively.  All the plots within
$0 \leq d \leq 0.5a$ overlap. The open squares and
the filled circles overlap for $0.5a < d \leq 1.0a$.
There are not local magnetic moments when the total
magnetization in a layer is zero.  This is the 
case of $0 \leq d \leq 0.5a$.  In this region,
all of the carbon atoms between layers interact
strongly.  The hopping interactions tend to enhance
the itinerancy of electrons in the direction perpendicular
to the layers.  This enhancement of the itinerancy
suppresses the magnetic orderings.  On the other
hand, a certain magnitude of the local magnetic
moments exist for larger $d$ (near the A-B stacking).
The magnetic polarization is negative along the 
edge A-A', and it is positive along the edges
B-C and B'-C' in the first layer of Fig. 1 (a).
The value of the polarization at the inner atoms 
apart from the edges becomes smaller.  The carbon 
atoms near the edges have weak interactions
with the other atoms of neighboring layers.
Such the property gives rise to the weak itinerant
character of electrons, and this is the main
origin of the appearance of the local magnetic
polarization at the edge atoms.  The first
order phase transition between the small and
large $d$ regions characterizes the qualitative
difference of the magnetic properties.

In the band calculations of the stacked nanographite
ribbons [15], the strong hybridization between edge 
states occurs in the A-A stacking case.  Such the 
hybridization is weak in the A-B stacking case.  The 
strong localization of wavefunctions at the edge 
carbon sites persists in the band calculations for 
systems with the A-B stacking [15], and this property 
agrees with the present result of the appearance
of local magnetic polarization near the A-B stacking.

\section{Density of states}

In this section, we discuss the local density of states
at the edge sites.  The wavefunctions of electrons with
up and down spins are projected on the edge sites which
are labeled in Fig. 1 (a).  The local density of states 
is reported together with the total density of states.

Figure 5 reports the density of states per layer while 
$d$ is varied: $d=0$ for Fig. 5 (a), $d=0.5a$ for Fig. 5 (b), 
and $d=1.0a$ for Fig. 5 (c).  The other parameters are 
taken constant: $U=2.5t$ and $R=3.4$\AA.  The bold line 
shows the density of states over 24 carbon atoms per layer 
and per spin.  The thin and dashed lines indicate the 
density of states over the eight edge sites in a layer 
for the up and down spins, respectively.  We note that 
the thin lines and the dashed lines accord with each other
in Figs. 5 (a) and (b).  This is related with the 
property that magnetizations are not present.  On the other hand,
the up and down splitting typical to the antiferromagnetism 
is seen in Fig. 5 (c).  Because the number of edge sites is 
one third of that of the total carbon atoms in the 
nanographite {\bf a}, the areas between the lines and
the horizontal axis have such the relative ratios.
In one dimensional graphite ribbons [4-6], there appears
a strong peak due to the localized edge states at the
Fermi energy.  This is seen in the non-interacting case.
With interactions taken into account, such the edge
states split into bonding (occupied) and antibonding
(unoccupied) states.  This fact will be one of the reasons
why such the strong peak is not observed in Fig. 5.
Also, in the present case, the edge sites do not make 
a one dimensional lattice and each layer has a finite
spatial dimension.  Such the difference will be the
second reason of the absence of the strong peak.

In the experiments of nanographite, for example in [14], 
the stacking patterns have not been observed directly.  
However, our theoretical results clearly show that 
the A-B type stacking is favorable in nanometer
size graphite systems in order that the exotic magnetic 
properties [1-3] are to be observed experimentally.

\section{Summary}

In summary, we have theoretically investigated the stacking 
effects in the zigzag nanographite sheets.  
We have found a remarkable difference in the magnetic 
properties between the simple A-A and A-B type stackings. 
For the simple stacking, there are not magnetic solutions.
For the A-B stacking, we have found the antiferromagnetic solutions 
for strong onsite repulsions.  The local magnetic moments 
tend to exist at the edge sites in each layer due to the 
large amplitude of wavefunctions at these sites.  We have 
studied the continuous changes between the A-A and A-B stackings,
and have found that the magnetism between the two stackings
experiences the first order phase transition.  The effect
of the interlayer distance has been discussed.
Therefore, the A-B type stacking is favorable in order that the
exotic magnetism is observed in nanographite systems.

\mbox{}

\begin{flushleft}
{\bf Acknowledgements}
\end{flushleft}

\noindent
The author is grateful for interesting discussion with
T. Enoki, N. Kawatsu, T. Ohshima, Y. Miyamoto, K. Kusakabe, 
K. Nakada, K. Wakabayashi, and M. Igami.  Useful discussion 
with the members of Condensed Matter Theory Group
(\verb+http://www.etl.go.jp/+\~{}\verb+theory/+),
Electrotechnical Laboratory is acknowledged, too.

\pagebreak
\begin{flushleft}
{\bf References}
\end{flushleft}

\noindent
$[1]$ Y. Shibayama, H. Sato, T. Enoki, and M. Endo, 
Phys. Rev. Lett. {\bf 84}, 1744 (2000).\\
$[2]$ N. Kobayashi, T. Enoki, C. Ishii, K. Kaneko, and M. Endo,
J. Chem. Phys. {\bf 109}, 1983 (1998).\\
$[3]$ N. Kawatsu, H. Sato, T. Enoki, M. Endo, 
R. Kobori, S. Maruyama, and K. Kaneko,
Meeting Abstracts of the Physical Society of Japan
{\bf 55} Issue 1, 717 (2000).\\
$[4]$ M. Fujita, K. Wakabayashi, K. Nakada, and K. Kusakabe,
J. Phys. Soc. Jpn. {\bf 65}, 1920 (1996).\\
$[5]$ M. Fujita, M. Igami, and K. Nakada,
J. Phys. Soc. Jpn. {\bf 66}, 1864 (1997).\\
$[6]$ K. Nakada, M. Fujita, G. Dresselhaus, and M. S. Dresselhaus,
Phys. Rev. B {\bf 54}, 17954 (1996).\\
$[7]$ R. Saito, M. Yagi, T. Kimura, G. Dresselhaus, and
M. S. Dresselhaus, J. Phys. Chem. Solids {\bf 60},
715 (1999).\\
$[8]$ M. S. Dresselhaus and G. Dresselhaus,
Adv. Phys. {\bf 30}, 139 (1981).\\
$[9]$ P. S. Surj\'{a}n, Int. J. Quantum Chem.,
{\bf 63}, 425 (1997).\\
$[10]$ R. Saito, G. Dresselhaus, and M. S. Dresselhaus,
J. Appl. Phys. {\bf 72}, 494 (1993).\\
$[11]$ S. Stafstr\"{o}m, Phys. Rev. B {\bf 43},
9158 (1991).\\
$[12]$ D. Baeriswyl, and K. Maki, Phys. Rev. B
{\bf 28}, 2068 (1983).\\
$[13]$ K. Harigaya, Phys. Rev. B {\bf 53}, R4197 (1996).\\
$[14]$ T. Suzuki and K. Kaneko, Carbon {\bf 26},
743 (1988).\\
$[15]$ Y. Miyamoto, K. Nakada, and M. Fujita,
Phys. Rev. B {\bf 59}, 9858 (1999).\\

\pagebreak
\begin{flushleft}
{\bf Figure Captions}
\end{flushleft}

\mbox{}

\noindent
Fig. 1. Stacked nanographite with zigzag edges.
The bold and thin lines show the first and second
layers, respectively.  The stacking is the A-B type 
in (a) (nanographite {\bf a}), and it is the simple 
A-A type in (b) (nanographite {\bf b}).  There are 
24 carbon atoms in one layer.  The circles in (a)
show sites where the interlayer distance between 
carbon atoms is shortest.  The edge sites, A (B, C) 
and A' (B', C'), are symmetrically equivalent, 
respectively.

\mbox{}

\noindent
Fig. 2.  The magnitude of the total magnetic moment
per layer as a function of $d$ and $U$.  The interlayer
distance is $R=3.4$\AA.  The values of $U$ are $U=2.1t$ 
(filled squares), $2.2t$ (open squares), $2.3t$ (filled circles),
$2.4t$ (open circles), $2.5t$ (filled triangles),
and $2.6t$ (open triangles), respectively. 
All the plots within $0 \leq d \leq 0.5d$ overlap,
so only the squares are seen.

\mbox{}

\noindent
Fig. 3.  The magnitude of the total magnetic moment
per layer as a function of $d$ and $R$.  The strength
of the Coulomb interaction is $U=2.5t$.  The values 
of $R$ are $R=3.4$\AA (filled squares), 3.2\AA
(open squares), and 3.0\AA (filled circles), 
respectively.  All the plots within $0 \leq d \leq 0.5d$ 
overlap, so only the filled squares are seen.

\mbox{}

\noindent
Fig. 4.  Local magnetic moment at the edge sites
A, B, and C, as a function of $d$.  The parameters are 
$U=2.5t$ and $R=3.4$\AA.  The filled squares, open 
squares, and filled circles show the results at sites 
A, B, and C, respectively.  All the plots within
$0 \leq d \leq 0.5a$ overlap. The open squares and
the filled circles overlap for $0.5a < d \leq 1.0a$.
Therefore, the number of plots seems smaller.

\mbox{}

\noindent
Fig. 5.  Density of states per layer while $d$ is varied:
$d=0$ for (a), $d=0.5a$ for (b), and $d=1.0a$ for (c).
The other parameters are $U=2.5t$ and $R=3.4$\AA.
The bold line shows the density of states over 24 carbon 
atoms per layer and per spin.  The thin and dashed lines 
indicate the density of states over the eight edge sites
in a layer for the up and down spins, respectively.
We note that the thin lines and the dashed lines 
overlap in Figs. (a) and (b).

\end{document}